\documentclass[preprint,11pt]{elsarticle}
\usepackage{setspace}

\usepackage{amsmath}
\usepackage[utf8]{inputenc} 
\usepackage[T1]{fontenc}    
\usepackage{hyperref}       
\usepackage{url}            
\usepackage{booktabs}       
\usepackage{amsfonts}       
\usepackage{nicefrac}       
\usepackage{microtype}      
\usepackage{lipsum}
\usepackage{fancyhdr}       
\usepackage{graphicx}       
\graphicspath{{media/}}     

\usepackage{natbib}
\usepackage{tabularx}
\usepackage[flushleft]{threeparttable}
\usepackage{tcolorbox}
\usepackage{algorithm}
\usepackage{algorithmic}
\usepackage{float}
\usepackage{multirow}

\begin{document}

\begin{frontmatter}

\title{CrossTraffic: An Open-Source Framework for Reproducible and Executable Transportation Analysis and Knowledge Management}

\author{Rei Tamaru and Bin Ran} 

\affiliation{organization={University of Wisconsin-Madison},
            addressline={1415 Engineering Dr.},
            city={Madison},
            postcode={53706},
            state={Wisconsin},
            country={United States}}

\begin{abstract}
Transportation engineering often relies on technical manuals and analytical tools for planning, design, and operations. However, the dissemination and management of these methodologies, such as those defined in the Highway Capacity Manual (HCM), remain fragmented. Computational procedures are often embedded within proprietary tools, updates are inconsistently propagated across platforms, and knowledge transfer is limited. These challenges hinder reproducibility, interoperability, and collaborative advancement in transportation analysis.

This paper introduces CrossTraffic, an open-source framework that treats transportation methodologies and regulatory knowledge as continuously deployable and verifiable software infrastructure. CrossTraffic provides an executable computational core for transportation analysis with cross-platform access through standardized interfaces. An ontology-driven knowledge graph encodes engineering rules and provenance and serves as a semantic validation layer for analytical workflows. A conversational interface further connects large language models to this validated execution environment through structured tool invocation, enabling natural-language access while preventing procedurally invalid analyses.

Experimental results show that knowledge-graph-constrained execution substantially improves numerical accuracy and methodological fidelity compared with context-only approaches, achieving near-zero numerical error (MAE $<0.50$) across multiple large language models and perfect detection of invalid analytical inputs in stress testing (F1~=~1.0). Its modular architecture supports the integration of additional transportation manuals and research models, providing a foundation for an open and collaborative transportation science ecosystem with a reproducible computational core. The system implementation is publicly available at \url{https://github.com/crosstraffic}.
\end{abstract}


\end{frontmatter}

\section{Introduction}
Designing safe and efficient transportation systems has long been a central tenet of transportation engineering, supported by a shared body of knowledge, from the empirical equations of the Highway Capacity Manual (HCM) to the regulatory definitions of the American Association of State Highway and Transportation Officials (AASHTO). However, the current ecosystem for managing this transportation knowledge is largely closed and fragmented. Critical methodologies are typically locked within proprietary, black-box software tools or unstructured documents (PDFs) \cite{oman2009}. This lack of machine-interpretable representations indicates that transportation knowledge is often dispersed and difficult to verify, limiting opportunities for reuse, collaborative innovation, and rigorous reproducibility of analytical results \cite{kleinsteuber2024}.

In response to these challenges, earlier efforts to improve knowledge sharing have emphasized the concept of Transportation Knowledge Networks -- decentralized infrastructures that connect information providers and users across agencies and jurisdictions \cite{oman2009, jenks2015}. While such initiatives improved access to documentation and facilitated technology transfer, computational interoperability for transportation methodologies remains elusive. As a result, the absence of a unified, machine-readable representation of transportation methodologies continues to impede automated validation, cross-tool integration, and reliable deployment of simulation-driven decision support systems in Advanced Intelligent Transportation Systems (ITS) \cite{harrison2025}.

To bridge this gap, recent research has increasingly adopted ontology-driven architectures in different fields \cite{bai2025}. By representing domain knowledge as a structured Knowledge Graph (KG), this enables the transformation of static manuals into \textit{executable standards} that constrain and validate computation workflows. This approach is particularly critical in the era of Generative AI, where Large Language Models (LLMs) offer powerful interfaces but lack the inherent reliability to perform safety-critical engineering without structured guardrails \cite{bubeck2023}. However, a KG alone is insufficient, and it must be embedded within an open and auditable software ecosystem that supports community maintenance and review of the encoded logic.

To realize this paradigm, we introduce \textbf{CrossTraffic}, an open-source knowledge management system (KMS) for representing and executing transportation engineering logic. In CrossTraffic, regulatory constraints, analytical procedures, and validation rules are encoded as auditable and reusable digital artifacts, rather than embedded within monolithic software applications. The system separates semantic validation, computational execution, and user interaction into independent components, enabling modular evolution while preserving analytical correctness and reproducibility.

From its inception, CrossTraffic has evolved beyond a simple calculation tool into a robust KMS. This paper documents the framework's architecture, its ontological underpinnings, and its validation as a platform for modern transportation systems. Our primary contributions are:

\begin{enumerate}
    \item \textbf{A Framework for Open Knowledge Maintenance:} We propose a modular architecture that decouples regulatory logic from computational execution, significantly lowering the barrier for community contribution and maintenance.
    
    \item \textbf{Ontological Standardization of Terminology:} We demonstrate how the domain-specific KG effectively manages the complex relationships between transportation terminologies.
    
    \item \textbf{Cross-Platform Reproducibility \& Reliability:} By centralizing knowledge in a computational core, our model-agnostic fidelity experiments show that integrating this structured knowledge with LLMs reduces computational error by \textbf{$> 94\%$}, solving the black-box reproducibility crisis inherent in generative AI.
\end{enumerate}

\section{Related Works}
\subsection{Knowledge Management for Open Transportation Science}
Recent work in knowledge management has proposed structured, open, and collaborative approaches for organizing transportation information \cite{pangaribuan2024}. Effective practices in the domain involve capturing, curating, and sharing expertise across stakeholders to support informed decision-making and enable organizational learning \cite{ettore2018, olan2022, jarrahi2023}. However, transportation agencies face significant barriers to achieving this vision. For instance, siloed information systems often exist in state departments and municipal agencies, leading to incompatible systems for traffic data, manuals, and analytical tools that limit inter-agency interoperability \cite{jenks2015}. Besides, proprietary software encapsulates fundamental methodologies, limiting transparent access and collaborative refinement of analytical procedures \cite{weerakkody2021}. To address these challenges, knowledge management research has proposed curation-oriented models, such as the Digital Curation Center model \cite{higgins2008} and the framework of \citet{irani2023}, which emphasize systematic description, representation, preservation, and reuse of information resources.

Building on these knowledge management and curation efforts, transportation practice increasingly relies on data-driven and simulation-based decision support systems, which require the integration of heterogeneous data sources with formal analytical models \cite{chowdhury2024}. In parallel, digital twin initiatives have further expanded the use of real-time data coupled with microscopic simulation for operational planning and system optimization \cite{kusic2023}. Creating a middleware system for cross-platform interoperability, in turn, is a prerequisite for a cohesive and robust testbed for smart city applications \cite{goumopoulos2024}. At the organizational level, sustained analytical practice also depends on effective knowledge sharing and cross-disciplinary collaboration in safety, mobility, and infrastructure planning \cite{kleinsteuber2024, irfan2022}. Despite these advances, existing infrastructures primarily support data and document exchange rather than executable representations of analytical procedures. Consequently, the absence of a unified, machine-executable representation of codified transportation knowledge continues to limit reproducibility and scalable interoperability for traffic operations and simulation-based decision support \cite{santana2017, riehl2025}.

\subsection{Semantic Intelligence and Decision Support Layers}

In this paper, we use the term \emph{semantic intelligence} to refer to language-centered interfaces and representation learning techniques that support interpretation, retrieval, and explanation of transportation knowledge through natural language interaction.

As transportation systems grow increasingly complex, there is increasing demand for artificial intelligence-based decision support tools to help practitioners navigate and query large bodies of domain knowledge \cite{katsumi2018, zhang2025}. Recent advances in LLMs and retrieval-augmented generation (RAG) \cite{lewis2020} have enabled natural language interfaces for transportation knowledge access and decision support \cite{tupayachi2024, ye2025}. These systems allow users to pose queries in natural language and retrieve relevant methodological descriptions, reports, or documentation, improving accessibility and usability for non-expert users \cite{lu2025}.

In the transportation domain, such language-driven semantic interfaces have been explored for a range of applications, including traffic management support \cite{masri2025}, interaction with digital twin platforms \cite{yang2024}, and adaptive control of connected and automated vehicles \cite{cui2024}. Despite these advances, LLM-based semantic layers remain fundamentally probabilistic and operate primarily over unstructured or weakly structured text. As a result, existing studies report persistent limitations in analytical reliability, completeness of domain coverage, and standardization of tool integration \cite{nie2025}. 

A central challenge is that most current LLM- and RAG-based systems operate independently of authoritative engineering manuals and regulatory procedures. This separation makes it difficult to guarantee that retrieved or generated analytical workflows comply with established standards, and can lead to semantic hallucinations and procedural inconsistencies when models infer or omit required steps.

\subsection{Ontology-Based Knowledge Management in Transportation}

In contrast to language-centered semantic intelligence, ontology-based knowledge management focuses on formally representing domain concepts, rules, and relationships in machine-interpretable structures that support deterministic reasoning.

Ontology-based Knowledge Management Systems (OKMS) provide a formal and explicit specification of shared conceptualizations, enabling software systems to reason about entities, attributes, and relationships \cite{gruber1993, studer1998, hogan2021}. As reviewed by \citet{mora2022}, OKMS offers a structured framework for transforming tacit domain knowledge into explicit, machine-executable logic. By leveraging W3C standards such as the Resource Description Framework and the Web Ontology Language \cite{McBride2004}, these systems support formal reasoning and structured integration of heterogeneous datasets across multiple domains. Large-scale infrastructures such as KnowWhereGraph \cite{janowicz2022} illustrate one successful application of ontology-based integration in the geospatial and infrastructure context.

Within the transportation domain, ontologies have been successfully applied to support data interoperability and semantic integration. For example, \citet{fernandez2016} proposed an ontology-driven architecture based on the Semantic Sensor Network ontology to interpret real-time traffic situations from heterogeneous sensor streams. \citet{zhang2025} employed KGs to represent mobility-related entities and activities, combining rule-based and neural models to infer semantic trip purposes from numerical data. Similarly, \citet{gan2025} demonstrated the use of KGs for underground infrastructure management and applied entity disambiguation techniques to consolidate redundant data records.

Despite these successes, a significant gap remains in the use of ontologies for normative and regulatory reasoning in transportation engineering. Existing works primarily support descriptive interoperability, such as data integration and record consolidation (e.g., \cite{gan2025}), or semantic decision support in other application domains (e.g., \cite{moradi2013}). However, these approaches do not establish an explicit connection between ontology-based knowledge representations and the execution of transportation engineering analyses and regulatory validation procedures. This work addresses this gap by introducing an executable regulatory ontology that links formal domain concepts and rules to concrete analytical procedures that verify multi-source regulatory and methodological constraints prior to computational execution.

\section{System Architecture} \label{sec:architecture}
\begin{figure}[h]
  \centering
  \includegraphics[width=\textwidth]{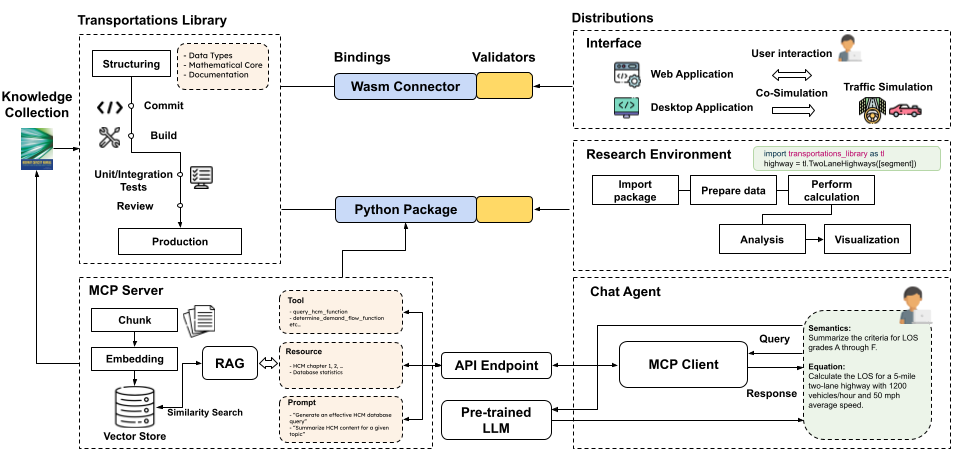}
  \caption{Modular architecture of the CrossTraffic platform. The system is stratified into three integrated layers: the Computational Core (Transportations Library) for authoritative calculation, Middleware for providing bindings, and the Interaction Ecosystem for distributions of capabilities across multi-platforms. An ontology-driven semantic validation layer operates across these components to validate analytical requests and parameters.} \label{fig:architecture}
\end{figure}

To support the continuous deployment and advancement of transportation knowledge, this paper introduces CrossTraffic, a unified, open-source framework designed for systematic knowledge management and computational integration. CrossTraffic provides a structured approach that decouples the definition of rules from their execution.

As illustrated in Figure \ref{fig:architecture}, the framework consists of three connected components: \textbf{(1) The Computational Core}, a high-performance Rust library that serves as the authoritative repository for transportation equations and data structures. \textbf{(2) The Middleware Layer}, which exposes the core to multiple programming environments through Python bindings and WebAssembly (WASM). \textbf{(3) The Interaction Ecosystem}, which delivers user-facing applications such as web calculators, desktop clients, and intelligent agents. Across these components, CrossTraffic incorporates an ontology-driven semantic validator that inspects analytical requests and parameters before execution. The design and operation of this validation layer are described in detail in Section~\ref{sec:methodology}.

\subsection{Computational Core: Transportations Library} \label{sec:core}
The \texttt{Transportations Library} forms the foundational layer of CrossTraffic and provides the authoritative implementation of transportation analysis procedures. The library is implemented with a strongly typed and memory-safe programming model \cite{matsakis2014}, enabling compile-time enforcement of parameter structure and dimensional consistency. Rather than treating engineering inputs as unconstrained numeric values, domain parameters are represented as explicit data types.

Each analytical function and equation from established transportation manuals is encapsulated as a modular component with well-defined interfaces. To maintain the integrity of this source of truth, the library adheres to a Test-Driven Development paradigm. A dedicated Continuous Integration and Continuous Development pipeline runs unit tests against manual examples and regression tests against previous builds on every commit, ensuring that the equations remain consistent over time.

\subsection{Middleware: Python Bindings and WASM Wrappers} \label{sec:middleware}

To make the Transportations Library broadly accessible across diverse computing environments, CrossTraffic incorporates a middleware layer that compiles the Rust kernel into portable targets. This layer exposes the library’s computational capabilities through Python bindings and WebAssembly (WASM) \cite{webassembly2022} wrappers.

The Python interface, engineered using PyO3 \cite{pyo3}, allows researchers and practitioners to seamlessly integrate the library's authoritative calculations directly into their data analysis and modeling workflows, leading to high-throughput analysis without reimplementing the underlying equations. For deployment in web and other cross-platform applications, CrossTraffic utilizes a WASM wrapper. This wrapper compiles the library into portable bytecode, facilitating several key advantages. This enables browser-based calculators to execute the exact same binary as the server-side analyses, guaranteeing consistent results across operating systems without native installation.

\subsection{Cross-Platform User Interfaces} \label{sec:interfaces}
\begin{figure}[h]
  \centering
  \includegraphics[width=0.9\textwidth]{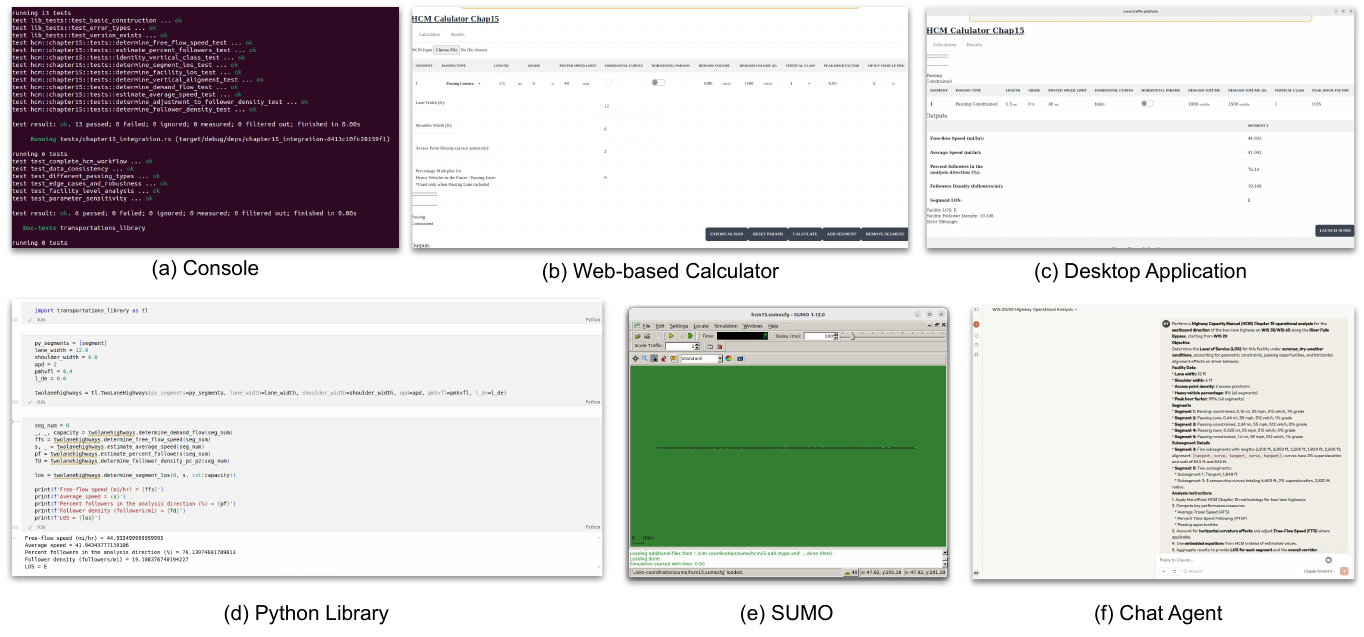}
  \caption{CrossTraffic deployed across multiple platforms. (a) Linux console. (b) Web-based calculator interface. (c) Desktop application packaged with Tauri. (d) Python library integration. (e) Hybrid co-simulation with SUMO. (f) MCP connection with the Claude Desktop. Each platform is validated using a consistent scenario setup to ensure reproducibility across environments.}\label{fig:platforms}
\end{figure}

CrossTraffic's user-facing layer provides practitioners and researchers direct access to transportation methodologies while preserving computational fidelity across diverse platforms (Figure \ref{fig:platforms}). The primary interface, the HCM Calculator, is distributed as both a Progressive Web App and a desktop client via Tauri.

Beyond static analysis, the framework facilitates dynamic validation through hybrid co-simulation. The calculator supports exporting inputs as SUMO configuration files \cite{lopez2018}. Utilizing the TraCI interface, the system can launch SUMO simulations directly from the desktop client. This establishes a direct connection between the analytical framework and a microscopic traffic simulator.

\subsubsection{Generative AI Integration (MCP)} \label{sec:mcp_integration}
The final component of the interaction ecosystem is the Model Context Protocol (MCP) server, which connects the computational core to generative AI agents. The MCP server exposes the functions of the Transportations Library as callable tools for LLMs. The MCP-based interface, therefore, combines language-based retrieval using RAG with structured tool invocation for authoritative computation and validation.

To support language-centered semantic interaction with transportation manuals and technical references, the interface adopts the RAG approach \cite{lewis2020}. Documentation (e.g., HCM chapters) is preprocessed and embedded into a vector database (e.g., ChromaDB) that leads to semantic similarity-based retrieval of relevant passages for user queries. The system then provides an interactive decision-support interface in which users may request explanations or analysis workflows (e.g., level-of-service computation or parameter interpretation), while all numerical results are generated through the verified computational core.

\section{The Ontology-Driven Architecture}
\label{sec:methodology}

\begin{figure}[ht]
    \centering
    \includegraphics[width=\textwidth]{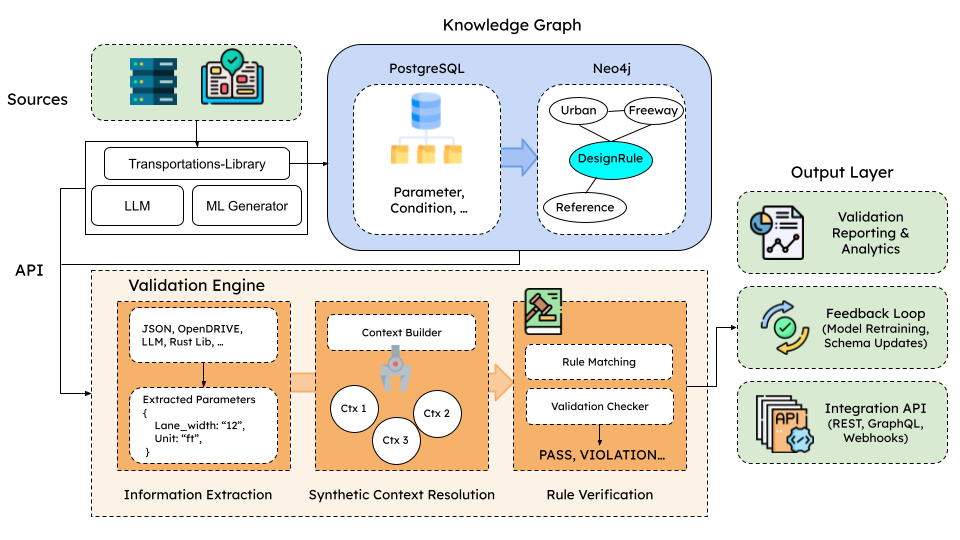}
    \caption{The validation execution pipeline. The pipeline converts heterogeneous design inputs into normalized parameters, resolves their regulatory context, retrieves applicable design rules from the KG, and evaluates all constraints through the Validation Engine. It then allows downstream computation and simulation to proceed. 
    }
    \label{fig:validator_pipeline}
\end{figure}

A fundamental challenge in modern transportation analytics lies not in numerical computation, but in the verification of contextual and regulatory validity. Standard software tools treat engineering inputs (e.g., lane width, grade) as generic floating-point numbers. In contrast, the CrossTraffic framework treats inputs as semantic entities constrained by formal regulatory and design rules derived from the HCM.

While Section \ref{sec:architecture} established the computational infrastructure, this section details the internal mechanics of the \textbf{Semantic Validator}. This component transforms the system from a passive data bridge into an active validation layer by validating all inputs against a formal KG external to the agent prior to execution.

As illustrated in Figure \ref{fig:validator_pipeline}, the validator enforces a domain-specific ontology. By structuring transportation knowledge as a formal graph $G=(\mathcal{V}, \mathcal{E})$, the system can execute graph queries to detect complex constraint violations (e.g., speed and radius incompatibility) that purely arithmetic validators would miss. The following subsections define the ontological structure, the formal constraint definitions, and the algorithmic execution pipeline.

\subsection{Domain-Specific Ontology Structure} \label{sec:ontology_structure}

Unlike traditional database schemas, which enforce data types, the Semantic Validator enforces \textit{engineering logic}. The KG schema is organized into five distinct node classes ($\mathcal{V}$) connected by semantic relationships ($\mathcal{E}$). This structure moves beyond simple look-up tables by explicitly encoding the dependencies between parameters, rules, and their authoritative sources.

\begin{figure}[h]
    \centering
    \includegraphics[width=0.8\textwidth]{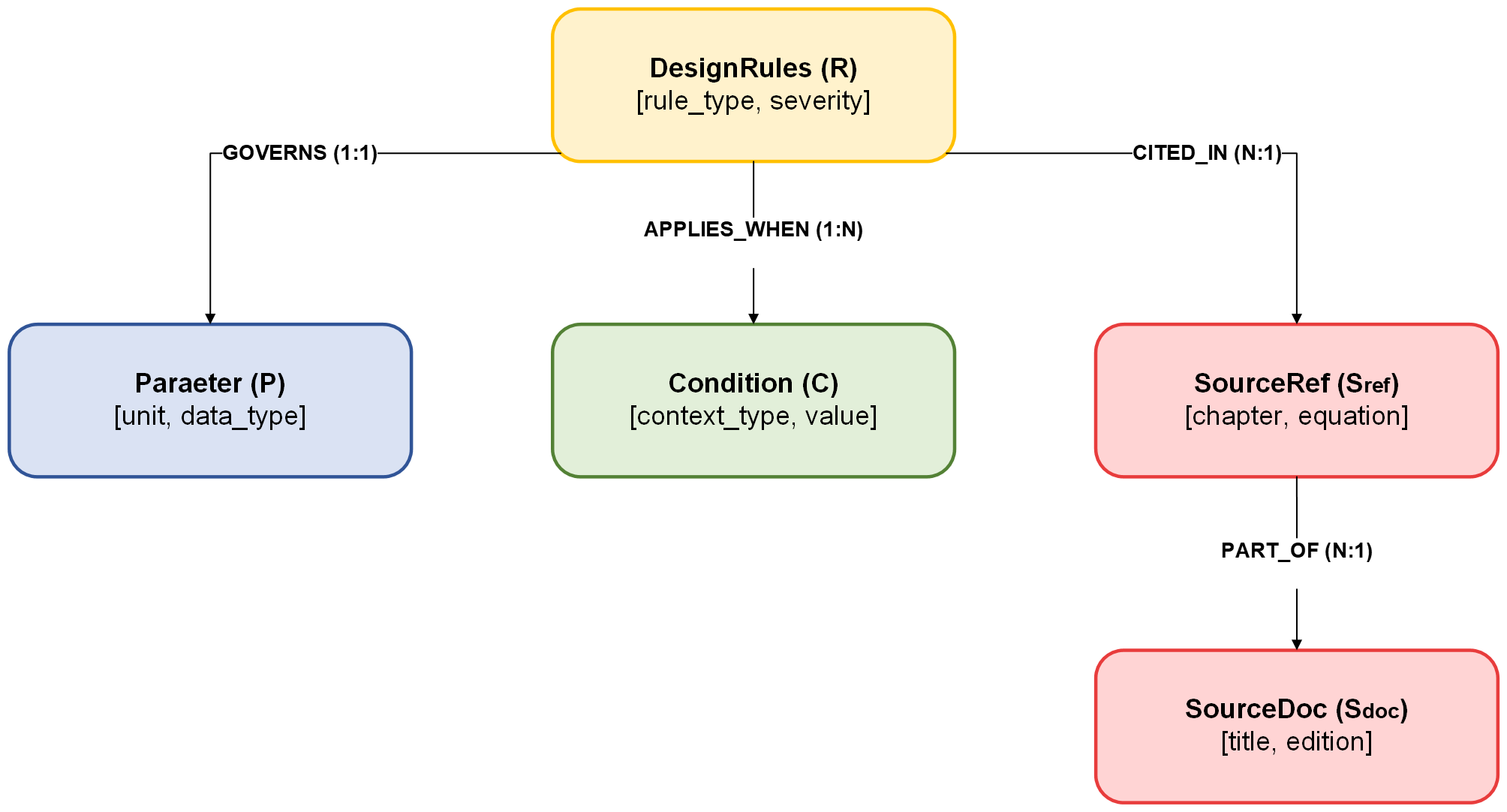} 
    \caption{The KG Ontology. Rules (Yellow) act as the central connectors, linking engineering Parameters (Blue) to their contextual Conditions (Green) and authoritative Sources (Red).}
    \label{fig:ontology_schema}
\end{figure}

\subsubsection{Node Entities ($\mathcal{V}$)}
The ontology defines the following entity types:
\begin{itemize}
    \item \textbf{Parameter ($P$):} Represents atomic engineering variables. Each node contains metadata that links it, via a foreign key, to the corresponding Rust \texttt{struct} field, ensuring type safety.
    \item \textbf{DesignRule ($R$):} Encodes the validation logic. These nodes possess a \texttt{rule\_type} attribute and a \texttt{severity} level, allowing for nuanced feedback.
    \item \textbf{Condition ($C$):} Represents the environmental context. These nodes act as logical gates, determining when specific rules are active.
    \item \textbf{Provenance ($S$):} Composed of \texttt{SourceDoc} (e.g., HCM 7th Ed.) and \texttt{SourceRef} (e.g., Chapter 15, Eq 15-3). These nodes ensure auditability by enabling every validation error to be traced to a specific legal or technical citation.
\end{itemize}

\subsubsection{Semantic Relationships ($\mathcal{E}$)}
The edges define the inferential capabilities of the system:
\begin{itemize}
    \item \texttt{(R)-[:VALIDATES]->(P)}: Links a logic rule to the parameter it governs.
    \item \texttt{(R)-[:REQUIRES]->(C)}: Establishes conditional logic. A rule node with this edge is only evaluated if the input scenario matches the connected Condition.
    \item \texttt{(P)-[:AFFECTS]->(P)}: Defines downstream dependencies, allowing the system to invalidate derived metrics when a root parameter changes.
    \item \texttt{(R)-[:CITED\_IN]->(S)}: Provides the audit trail, linking the mathematical logic to the text of the manual.
\end{itemize}

\subsubsection{Formal Constraint Definitions} \label{sec:formal_constraints}
Within this ontological structure, we populate the graph with specific \textbf{Semantic Validator constraints (SF-)} derived from the HCM (7th Edition) and AASHTO design standards. These constraints function as the instance layer of the ontology to ensure the logical robustness of the framework. We categorize them into three classes:

\begin{itemize}
    \item \textbf{Geometric Validity (SF-001, SF-002)}: Enforces dimensional limits calibrated to the empirical scope of the manual (e.g. $9 \leq \text{Lane Width} \leq 12\,\text{ft}$).
    \item \textbf{Operational Logic (SF-003, SF-004)}: Validates discrete categorical variables, ensuring that facility classifications (e.g., Passing Zones) align with the geometric context.
    \item \textbf{Physics-Based Safety (SF-005)}: A relational constraint enforcing the non-linear relationship between design speed ($V$) and minimum curve radius ($R_{\min}$), as defined by centripetal force and friction.
\end{itemize}

\begin{table}[ht]
\centering
\caption{Five Semantic Validator Constraints (Two-Lane Highway)}
\label{tab:constraints}
\renewcommand{\arraystretch}{1.2}
\begin{tabular}{@{}l l l l@{}}
    \toprule
    \textbf{ID} & \textbf{Parameter} & \textbf{Valid Range} & \textbf{Source} \\
    \midrule
    SF-001 & Lane Width & $9$--$12\,\text{ft}$ & HCM 7th Ed. \\
    SF-002 & Shoulder Width & $0$--$8\,\text{ft}$ & HCM 7th Ed. \\
    SF-003 & Horizontal Class & $0, 1, 2, 3, 4, 5$ & HCM 7th Ed. \\
    SF-004 & Passing Type & Constrained, Zone, Lane & HCM 7th Ed. \\
    SF-005 & Design Radius & $R \geq R_{min}(V_{design})$ & AASHTO Green Book \\
    \bottomrule
\end{tabular}
\end{table}

\subsection{The Validation Execution Pipeline}
\label{sec:validation_pipeline}
The operational core of the Semantic Validator is the validation pipeline. Unlike conventional input validation implemented directly in procedural code (e.g., \texttt{if x < 0}), CrossTraffic delegates validation to a graph-based rule evaluation process.

The pipeline executes four distinct phases for every computation request:
\begin{enumerate}
    \item \textbf{Semantic Mapping:} The incoming JSON payload is parsed, and keys are mapped to \texttt{Parameter} nodes ($P$) in the ontology. Unrecognized parameters are flagged immediately.
    \item \textbf{Context Resolution:} The system evaluates the global state (e.g., \texttt{FacilityType}) against \texttt{Condition} nodes ($C$) to determine the active rule set $R_{active} \subset R_{total}$.
    \item \textbf{Predicate Evaluation:} For each active rule $r \in R_{active}$, the system evaluates the logical predicate $f_r(p)$ against the input value $p$.
    \item \textbf{Enforcement:} If $\forall r, f_r(p) = \text{True}$, the validated inputs are serialized into Rust structs. If $\exists r, f_r(p) = \text{False}$, the pipeline halts and returns a structured \texttt{SemanticException}.
\end{enumerate}

Algorithm \ref{alg:semantic_firewall} details this logic. Note that the retrieval function \textsc{GetActiveRule} effectively pre-filters the graph, returning only those rules where the connected \texttt{Condition} node matches the input context $\mathbf{I}$, thereby optimizing the verification process.

\begin{algorithm}[H]
\caption{Semantic Validator Execution Logic}
\label{alg:semantic_firewall}
\begin{algorithmic}[1]
\REQUIRE User Input Vector $\mathbf{I}$, Knowledge Graph $G(\mathcal{V},\mathcal{E})$
\STATE $Errors \leftarrow \emptyset$
\FORALL{parameter $p \in \mathbf{I}$}
    \STATE $Node_p \leftarrow \text{FindParameterNode}(G, p.key)$
    \STATE $ActiveRules \leftarrow \text{GetActiveRules}(Node_p, \mathbf{I})$
    \FORALL{rule $r \in ActiveRules$}
        \IF{$\neg \text{Evaluate}(r, p.value)$}
            \STATE $Citation \leftarrow \text{TraverseSource}(r)$
            \STATE $Errors.\text{push}(\text{FormatError}(r, Citation))$
        \ENDIF
    \ENDFOR
\ENDFOR
\IF{$Errors \neq \emptyset$}
    \RETURN \text{Reject}(\textbf{400 Bad Request}, $Errors$)
\ELSE
    \RETURN \text{Proceed}(\textbf{200 OK}, \text{ExecuteRustCore($\mathbf{I}$)})
\ENDIF
\end{algorithmic}
\end{algorithm}

\section{Experimental Validation and Results} \label{sec:experiments}

To evaluate the operational feasibility and effectiveness of CrossTraffic as a unified knowledge management and execution framework, we conducted four validation studies. These studies assess the framework's ability to ensure operational consistency, logical robustness, digital-twin scalability, and generative-AI reliability.

\subsection{Ground Truth and Operational Consistency} \label{sec:gt}

We first established a ground-truth baseline to verify that the framework consistently executes HCM equations across heterogeneous user interfaces.

\subsubsection{Case Study: River Falls Bypass}

\begin{figure}[h]
  \centering
  \includegraphics[width=\textwidth]{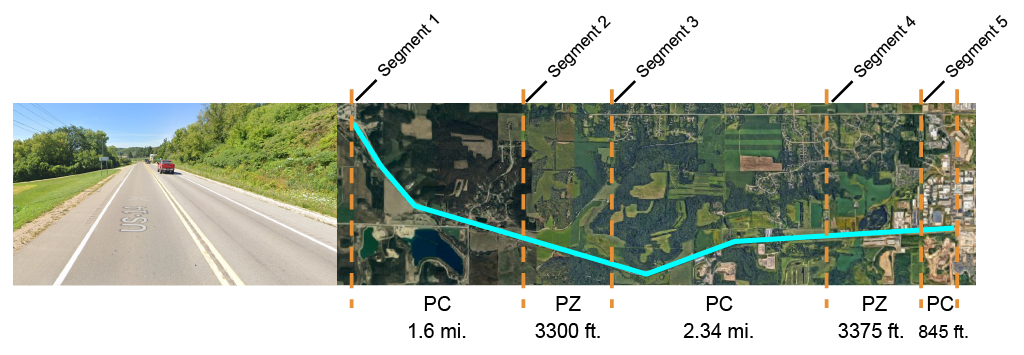}
  \caption{Case study visualization on WIS 35/WIS 65 along the River Falls Bypass}\label{fig:exp}
\end{figure}

We selected a real-world corridor: WIS~35/WIS~65 along the River Falls Bypass (Wisconsin), characterized by moderate geometric complexity and varying passing opportunities (Figure~\ref{fig:exp}). The analysis follows HCM Chapter~15 guidelines and considers five consecutive segments with mixed passing-constrained and passing zones, including horizontal curves with superelevation adjustments. 

\subsubsection{Inference Consistency Verification} \label{sec:consistency}

To evaluate cross-interface consistency and reproducibility, we performed the same operational analysis using both the deterministic Web Calculator (WASM) and the LLM-based semantic interface integrated through MCP.

\begin{table}[!ht]
\centering
\caption{Segment-Level and Overall LOS Comparison between Web Calculator (WASM) and LLM Interface} \label{tab:segment_wasm_vs_llm}
\resizebox{\textwidth}{!}{
    \begin{tabular}{ccccccccc}
    \toprule
    \multirow{2}{*}{\textbf{Segment}} & 
    \multicolumn{4}{c}{\textbf{Web Calculator (WASM)}} & 
    \multicolumn{4}{c}{\textbf{LLM Interface}} \\
     & AS (mph) & PF (\%) & FD (fol/mi) & LOS & AS (mph) & PF (\%) & FD (fol/mi) & LOS \\
    \midrule
    1 & 59.17 & 59.36 & 5.41 & C & 59.33 & 53.8 & 5.35 & C \\
    2 & 59.90 & 47.98 & 4.32 & C & 54.14 & 43.0 & 4.90 & C \\
    3 & 56.70 & 56.59 & 5.38 & C & 56.56 & 51.3 & 5.15 & C \\
    4 & 56.90 & 48.04 & 4.32 & C & 54.19 & 43.0 & 4.92 & C \\
    5 & 56.76 & 55.10 & 5.23 & C & 55.07 & 50.0 & 5.02 & C \\
    \midrule
    \textbf{Overall} & & & \textbf{5.09} & \textbf{C} & & & \textbf{5.05} & \textbf{C} \\
    \bottomrule
    \end{tabular}
}
\end{table}

As summarized in Table~\ref{tab:segment_wasm_vs_llm}, the results confirm the single source-of-truth architecture. Despite fundamentally different interaction modalities, both interfaces produced an identical overall level of service (LOS C). Minor discrepancies in average speed (AS) and follower density (DF)(within $0.05$) are attributed to floating-point rounding differences between the browser-based runtime and the server-side execution environment. These results confirm that the shared computational core and semantic interface produce consistent analytical results across the ecosystem.

\subsection{Logical Robustness and Constraint Enforcement} \label{sec:logical}

To quantify the regulatory capacity of the semantic validator, we conducted a stress test using a stochastic adversarial generator. Instead of relying on standard random testing, this generator employs boundary value analysis to target the specific edges of validity where software errors typically occur.

The generator produced $N=1000$ synthetic test vectors, including boundary attacks (e.g., $12.01$~ft lanes) and combinatorial conflicts (e.g., high speed combined with tight radius). The resulting dataset contained both compliant designs and diverse violations, including negative dimensions, excessive grades, and physics-incompatible speed-radius pairs.

\begin{table}[ht]
\centering
\caption{Confusion Matrix of Semantic Validator Stress Test} \label{tab:firewall}
\resizebox{\textwidth}{!}{
    \begin{tabular}{lcl}
        \toprule
        Metric & Value & Implications for Engineering Safety \\
        \midrule
        True Positives & 740 & The system effectively stops dangerous designs \\
        True Negatives & 260 & The system does not hinder valid workflows \\
        False Negatives & 0 & Critical safety success without illegal designs passed \\
        F1 Score & 1.00 & Perfect deterministic reliability \\
        \bottomrule
    \end{tabular}
}
\end{table}

As detailed in Table \ref{tab:firewall}, the framework rejected all invalid engineering inputs with no observed false positives. Notably, the median execution overhead was $0.002$ ms per check. This confirms that the KG-based validator can be integrated into high-frequency simulation loops without introducing measurable latency.

\subsection{Scalability and Heterogeneous Data Validation} \label{sec:scalability}

To verify scalability, we expanded the scope to include industry-standard digital-twin assets. We utilized the full OpenDRIVE asset suite from the CARLA Simulator (Town01 through Town07) \footnote{CARLA Assets on GitHub: \url{https://github.com/carla-simulator/opendrive-test-files/tree/master/OpenDrive}}. This dataset comprises $1,347$ road segments and $14,846$ parameter checks.

\begin{table}[ht]
\centering
\caption{Digital Twin Validation Results: CARLA Asset Suite ($N=7$ Towns)}
\label{tab:opendrive_scale}
\begin{tabular}{@{}l c c c c c@{}}
\toprule
\textbf{Asset ID} & \textbf{Roads} & \textbf{Params} & \textbf{Valid} & \textbf{Invalid} & \textbf{Pass Rate} \\
\midrule
Town01 (Urban) & 98 & 715 & 409 & 306 & 57.20\% \\
Town02 (Urban) & 68 & 824 & 452 & 372 & 54.85\% \\
\midrule
Town03 (Mixed) & 279 & 2,440 & 2,094 & 346 & 85.82\% \\
Town04 (Highway) & 242 & 2,708 & 2,433 & 275 & \textbf{89.84\%} \\
Town05 (Urban/Hwy) & 259 & 3,734 & 3,407 & 327 & \textbf{91.24\%} \\
Town06 (Highway) & 167 & 2,093 & 1,927 & 166 & \textbf{92.07\%} \\
Town07 (Rural) & 234 & 2,332 & 2,009 & 323 & 86.15\% \\
\midrule
\textbf{TOTAL} & \textbf{1,347} & \textbf{14,846} & \textbf{12,731} & \textbf{2,115} & \textbf{85.75\%} \\
\bottomrule
\end{tabular}
\end{table}

The validation results (Table~\ref{tab:opendrive_scale}) demonstrate the ability of the semantic validator to distinguish between geometrically valid road layouts and facilities that comply with U.S. design and operational standards.

\begin{itemize}
\item \textbf{Urban Non-Compliance:} Town01 and Town02 exhibited low pass rates ($\approx 55\%$) due to the use of $4.0\,\text{m}$ ($13.12\,\text{ft}$) lane widths to provide safety buffers for vehicle agents. While acceptable for robotics simulation, these dimensions violate HCM Constraint \texttt{SF-001} (Max $12\,\text{ft}$), which views excessive width as inducing unsafe speeds. Additionally, the urban grid layout introduces tight curve radii ($<200\,\text{ft}$) incompatible with the default $55\,\text{mph}$ highway classification logic, triggering \texttt{SF-005} (physics-based safety) violations.

\item \textbf{Highway Compliance:} Conversely, highway-focused maps (Town03-Town07) achieved high compliance ($>90\%$). The standard European highway lane width ($3.5\,\text{m} \approx 11.5\,\text{ft}$) falls perfectly within the HCM valid range ($9-12\,\text{ft}$), and the larger curve radii satisfy AASHTO safety envelopes.
\end{itemize}

This experiment demonstrates that the KG does not merely ingest data, but it evaluates it within its regulatory and operational context. For example, the system correctly identified that while Town01 represents a geometrically valid road layout, it does not constitute a valid two-lane highway facility under U.S. design standards.

\subsection{Model-Agnostic Performance Validation} \label{sec:model-agnostic}

Finally, we evaluated whether the framework mitigates hallucination and analytical inconsistency in generative-AI-based transportation analysis. We tasked four distinct agent architectures (GPT-5.2, Claude~4.5 Sonnet, Gemini~3.0 Pro, and Claude~4.5 Opus) with performing the River Falls case study.

In this study, \emph{context-based agents} refer to agents augmented only with uploaded PDF content, whereas \emph{MCP-integrated agents} combine RAG with tool invocation through the CrossTraffic MCP server and external semantic validation.

\subsubsection{Quantitative Convergence}

We measured the Mean Absolute Error (MAE) of computed traffic metrics (e.g., flow rate, density, speed) against the ground truth established in Section \ref{sec:gt}.

\begin{figure}[h]
    \centering
    \includegraphics[width=0.8\textwidth]{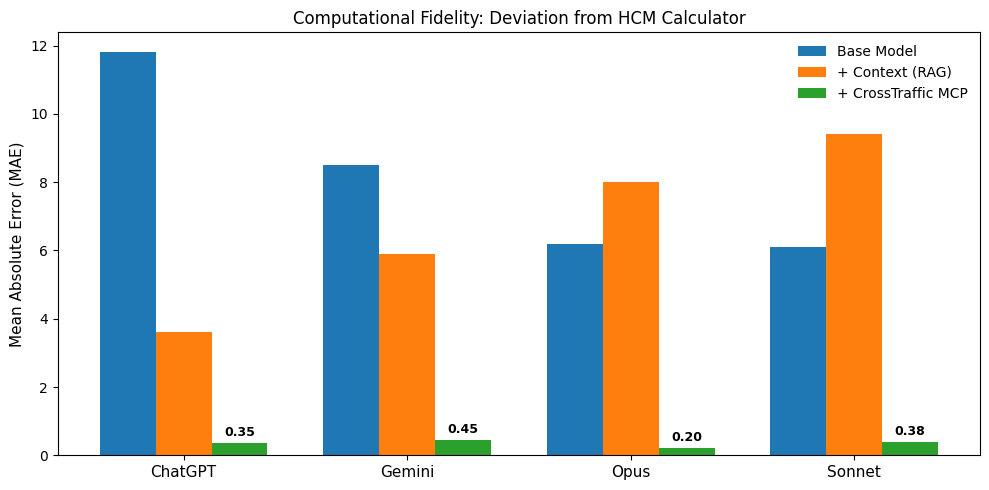}
    \caption{Impact of framework augmentation on computational error. Blue bars: base computational model with Rust-based calculator. Orange bars: PDF context augmentation for agents. Green bars: CrossTraffic MCP integration with agents.}
    \label{fig:mae_comparison}
\end{figure}

As illustrated in Figure~\ref{fig:mae_comparison}, PDF-based context augmentation exhibited unstable performance. In particular, providing an unstructured PDF context to Claude~4.5 Sonnet increased the error from 6.1 to 9.4, indicating difficulty in reliably interpreting procedural text. In contrast, all MCP-integrated agents converged to near-zero numerical error (MAE $< 0.50$), largely independent of the underlying language model.

\subsubsection{Qualitative Methodological Fidelity}

Beyond numerical accuracy, we assessed the reasoning quality of the engineering analysis. We defined a rubric (Table \ref{tab:criteria}) to score agent responses on a scale of 1 (Poor) to 3 (Excellent), providing with the same prompt.

\begin{table}[h]
\centering
\caption{Qualitative Assessment Criteria for Engineering Agents} \label{tab:criteria}
\footnotesize
    \begin{tabular}{@{}p{0.25\textwidth} p{0.70\textwidth}@{}}
    \toprule
    \textbf{Dimension} & \textbf{Definition of Excellent (Score 3)} \\
    \midrule
    \textbf{Methodological Fidelity} & Explicitly references specific HCM chapters/equations rather than generic rules of thumb. \\
    \textbf{Parameter Extraction} & Precisely maps user text to strongly-typed structs, capturing secondary details (e.g., superelevation). \\
    \textbf{Analytical Orchestration} & Provides a trace of the sequential analysis, linking specific library functions to results. \\
    \textbf{Hallucination Mitigation} & Strictly refuses to estimate missing data, requesting clarification instead of guessing. \\
    \textbf{Scientific Interpretability} & Provides key technical findings explaining why the facility performs at its current LOS. \\
    \bottomrule
    \end{tabular}
\end{table}

\begin{figure}[h]
\centering 
    \includegraphics[width=1.0\textwidth]{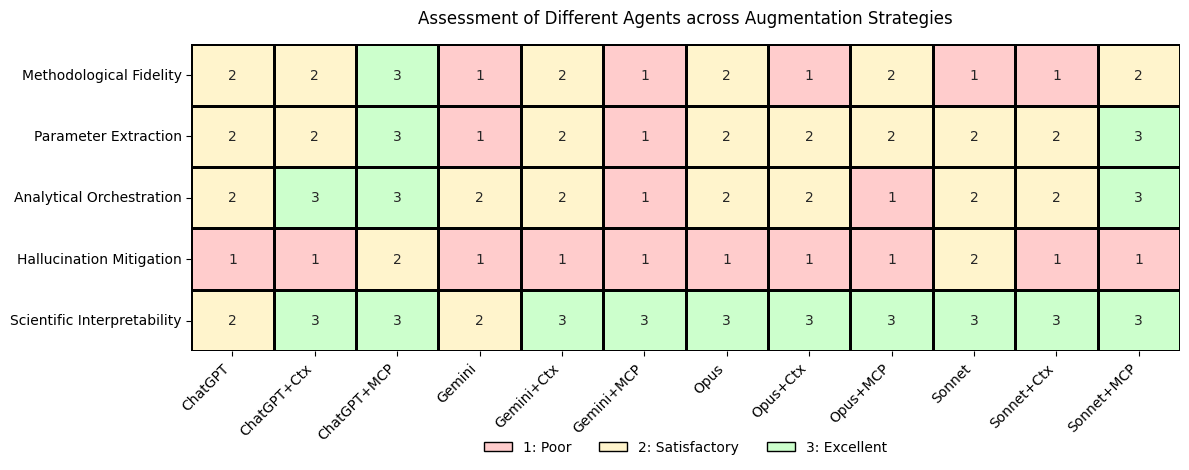}
    \caption{Qualitative scoring heatmap across multiple agents with different augmentation strategies. Lower numbers colored in red indicate poor performance on the metric, and higher numbers with green indicate better performance.}
    \label{fig:qualitative_heatmap}
\end{figure}

The heatmap results (Figure~\ref{fig:qualitative_heatmap}) reveal a clear performance contrast. While PDF-based context-augmented agents showed improved parameter extraction, they frequently failed in analytical orchestration, often producing numerical answers without an explicit computational trace. In contrast, MCP-integrated agents consistently achieved high scores in methodological fidelity by generating structured execution traces aligned with HCM procedures.

A trade-off was observed in contextual understanding. Because MCP-integrated agents prioritize structured tool execution, they occasionally underutilize implicit or underspecified user inputs compared with PDF-based context-only agents. In a small number of edge cases, strict schema requirements also encouraged models to infer missing parameters rather than explicitly requesting clarification. Nevertheless, scientific interpretability remained strong, particularly when agents were prompted to provide step-by-step validation of intermediate assumptions.

\section{Discussion}
The development of CrossTraffic represents a concrete step towards modernizing transportation analysis workflows by demonstrating the feasibility of transforming complex engineering manuals into a unified, continuously deployable knowledge management system. Beyond validating the technical feasibility of the proposed architecture, our findings reveal fundamental trade-offs in API design, agent orchestration, and organizational adoption that directly affect the practical deployment of AI-assisted engineering systems.

\subsection{API Granularity and the Agentic Trade-off}
Decomposing HCM procedures into reusable computational endpoints enables modularity and cross-platform deployment, but introduces an inherent trade-off between endpoint granularity and orchestration complexity. In our evaluation, baseline and MCP-integrated agents occasionally failed in analytical orchestration (Figure~\ref{fig:qualitative_heatmap}), primarily due to incorrect endpoint sequencing, missing intermediate steps, and inconsistent handling of intermediate states across multi-step workflows.

Effective use of AI agents further requires an \emph{extraction layer} that translates heterogeneous user inputs into structured parameters and \emph{evaluation layer} that presents results in a professionally interpretable form \cite{shankar2024, polak2024}. In the current implementation, parameter extraction and result presentation are handled through prompted LLM roles. While effective in our experiments, observed failure modes include implicit facility misclassification, unit ambiguity, and incorrect parameter binding when semantically similar variables coexist. These findings suggest that future agent controllers should employ dependency-aware planning over the API and KG, rather than relying on prompt-based endpoint chaining alone \cite{huang2024}.

\subsection{The Necessity of Semantic Guardrails}
A central empirical finding is that RAG does not guarantee engineering validity. As shown in Figure~\ref{fig:mae_comparison}, providing unstructured PDF context frequently degraded computational accuracy and increased numerical error, producing procedurally plausible but analytically invalid execution paths. We refer to such failures as \emph{semantic hallucinations}, where agents misapply equations or omit required analytical steps.

In contrast, MCP-integrated agents consistently converged to near-zero error across all evaluated foundation models (MAE $<0.50$), demonstrating that executable constraints and KG-based validation are essential for procedural reliability. While RAG remains valuable for document retrieval and conceptual explanation, it is insufficient for enforcing normative analytical workflows \cite{gupta2024}. Our results, therefore, support hybrid context-infused architectures in which LLMs interpret user intent while structured computational and ontological layers constrain and regulate permissible operations.

A corresponding limitation is a \emph{reading gap}: schema-constrained agents occasionally fail to exploit implicit or underspecified user context (Figure~\ref{fig:qualitative_heatmap}), because only explicitly provided or formally represented parameters can be passed to validated computational tools, while contextual assumptions expressed in natural language are not automatically translated into executable inputs.

\subsection{Open Source Significance for Transportation}
Finally, this work highlights persistent tension between proprietary black-box software and the scientific requirement for reproducibility \cite{wood2025}. Although the equations underlying most transportation analysis tools are publicly documented, their operational implementations are commonly embedded within closed-source systems. This practice creates substantial barriers for researchers and practitioners seeking to inspect, verify, or extend analytical methodologies.

However, CrossTraffic does not yet resolve the broader institutional and standardization challenges that limit open and interoperable transportation knowledge infrastructures. In particular, organizations responsible for developing official manuals must balance the protection of proprietary design processes and intellectual assets with the need for open and community-driven representations of foundational methodologies. Moreover, widely adopted, machine-readable standards for representing transportation equations, procedural rules, and regulatory logic are still lacking. As a result, large-scale interoperability and sustained knowledge exchange across tools and institutions remain open challenges beyond the scope of the current framework.

\section{Conclusion}
This paper presented CrossTraffic, a unified open-source software framework for maintaining, executing, and deploying transportation engineering knowledge. By combining a continuously tested, high-performance computational core with multi-language bindings and an ontology-driven semantic validation layer, CrossTraffic addresses persistent challenges in transportation practice, including fragmented implementations, limited interoperability across tools, and the lack of reproducible and auditable analytical workflows.

A central contribution of this work is the realization of an executable knowledge management architecture in which transportation procedures are represented as both validated computational functions and formal semantic constraints. This design establishes a single, authoritative source of analytical logic and enables consistent and reproducible results across heterogeneous interfaces, including web calculators, desktop applications, simulation pipelines, and conversational agents. Our experimental evaluation demonstrates that integrating large language models with this validated execution environment substantially improves computational fidelity and methodological correctness compared with context-only, retrieval-based approaches. Beyond numerical consistency, CrossTraffic demonstrates how an ontology-driven decision-support layer can safely bridge natural-language interaction and normative engineering workflows by validating regulatory and methodological constraints prior to execution.

Sustainable progress in this domain requires close collaboration among manual developers, software engineers, and transportation researchers. Establishing standardized and open representations of transportation equations and procedural logic can foster innovation while preserving regulatory accountability. 

\section{Acknowledgments}
The authors would like to acknowledge Jonathan Rehil for his valuable expert advice and insights during the development of this research. His contributions were instrumental in shaping the project.


\bibliographystyle{elsarticle-num-names}
\bibliography{references}

@techreport{harrison2025,
    author={Harrison, F. D. and Brown, C. and Admas, K. and Hall, T.},
    title={Knowledge management at state departments of transportation research roadmap},
    institution={National Academies of Sciences, Engineering, and Medicine},
    type={Research Report},
    number={1134},
    year={2025},
}

@techreport{oman2009,
    author={Oman, L. and Wilson, A. J. and Harrison, F. D.},
    title={Implementing transportation knowledge networks},
    institution={National Academies of Sciences, Engineering, and Medicine},
    type={Research Report},
    number={20-75},
    year={2009}
}

@article{irfan2022,
    title={Toward a resilient supply chain model: critical role of knowledge management and dynamic capabilities},
    author={Irfan, I. and Sumbal, M. S. U. K. and Khurshid, F. and Chan, F.TS},
    journal={Industrial management \& data systems},
    volume={122},
    number={5},
    pages={1153--1182},
    year={2022},
    publisher={Emerald Publishing Limited}
}

@article{matsakis2014,
    title={The Rust Language},
    author={Matsakis, N. D. and Felix S. K.},
    journal={Proceedings of the 2014 ACM SIGAda annual conference on High integrity language technology},
    year={2014},
}

@book{chowdhury2024,
    title={Data analytics for intelligent transportation systems},
    author={Chowdhury, M. and Dey, K. and Apon, A.},
    year={2024},
    publisher={Elsevier}
}

@misc{pyo3,
    author = {{PyO3 Project and Contributors}},
    license = {["Apache-2.0", "MIT"]},
    howpublished = {\url{https://github.com/PyO3/pyo3}},
    title = {{PyO3}},
    year = {2017--2025}
}

@techreport{webassembly2022,
    author={{WebAssembly Project and Contributors}},
    title = {{WebAssembly core specification}},
    version = {2.0},
    editor = {Rossberg, Andreas},
    date = {2022-04-19},
    year = {2022},
    institution = {{W3C}},
    url = {https://www.w3.org/TR/wasm-core-2/},
    langid = {english},
}

@techreport{jenks2015,
    author={Jenks, C. W. and Hedges, C. and Lemer, A. C. and Moore, S. and Delaney, E. P. and Hitchcock, S. E.},
    title = {A guide to agency-wide knowledge management for state departments of transportation},
    institution={National Academies of Sciences, Engineering, and Medicine},
    type={NCHRP Report},
    number={813},
    year = {2015}
}

@article{riehl2025,
    author={Riehl, K. and Kouvelas A. and Michail A. M.},
    title={Revisiting reproducibility in transportation simulation studies},
    journal={European Transport Research Review},
    volume={17},
    number={22},
    year={2025},
}

@article{kusic2023,
    title = {A digital twin in transportation: Real-time synergy of traffic data streams and simulation for virtualizing motorway dynamics},
    journal = {Advanced Engineering Informatics},
    volume = {55},
    pages = {101858},
    year = {2023},
    issn = {1474-0346},
    author = {K. Kušić and R. Schumann and E. Ivanjko},
}

@article{pangaribuan2024,
    author = {Pangaribuan, L. and Satrya, A.},
    year = {2024},
    month = {12},
    pages = {355-371},
    title = {The role of knowledge management, transformational leadership, and organizational commitment on employee performance: empirical study in public sector},
    volume = {17},
    journal = {Journal of Theory and Applied Management},
}

@article{olan2022,
    title = {Artificial intelligence and knowledge sharing: Contributing factors to organizational performance},
    journal = {Journal of Business Research},
    volume = {145},
    pages = {605-615},
    year = {2022},
    issn = {0148-2963},
    author = {Olan, F. and Arakpogun, O. E. and Suklan, J. and Nakpodia, F. and Damij, N. and Jayawickrama, U.},
}

@book{ettore2018,
    publisher={Springer},
    series={Knowledge Management and Organizational Learning},
    edition={None},
    booktitle={Emergent Knowledge Strategies},
    chapter={1},
    author={Bolisani, E. and Bratianu, C.},
    title={The elusive definition of knowledge},
    year={2018},
    month={March},
    pages={1-22},
    volume={None},
}

@article{jarrahi2023,
    title = {Artificial intelligence and knowledge management: A partnership between human and AI},
    journal = {Business Horizons},
    volume = {66},
    number = {1},
    pages = {87-99},
    year = {2023},
    issn = {0007-6813},
    author = {Jarrahi, M. H. and Askay, D. and Eshraghi, A. and Smith, P.},
}

@article{higgins2008,
    title={The DCC curation lifecycle model},
    author={Higgins, S.},
    journal={International Journal of Digital Curation},
    volume={3},
    number={1},
    pages={134-140},
    year={2008}
}

@article{weerakkody2021,
    author={Weerakkody, V. and Janssen, M. and El-Haddadeh, R.},
    title={The Resurgence of Business Process Re-Engineering in Public Sector Transformation Efforts: Exploring the Systemic Challenges and Unintended Consequences},
    journal={Inf Syst E-Bus Manage},
    volume={19},
    number={},
    pages={993-1014},
    year={2021}
}

@article{irani2023,
    title = {The impact of legacy systems on digital transformation in European public administration: Lesson learned from a multi case analysis},
    journal = {Government Information Quarterly},
    volume = {40},
    number = {1},
    pages = {101784},
    year = {2023},
    issn = {0740-624X},
    author = {Z. Irani and R. M. Abril and V. Weerakkody and A. Omar and U. Sivarajah},
}

@article{goumopoulos2024,
    author={Goumopoulos, C.},
    journal={IEEE Access}, 
    title={Smart city middleware: A survey and a conceptual framework}, 
    year={2024},
    volume={12},
    number={},
    pages={4015-4047},
}

@article{santana2017,
    author = {Santana, E. F. Z. and Chaves, A. P. and Gerosa, M. A. and Kon, F. and Milojicic, D. S.},
    title = {Software platforms for smart cities: Concepts, requirements, challenges, and a unified reference architecture},
    year = {2017},
    issue_date = {November 2018},
    publisher = {Association for Computing Machinery},
    address = {New York, NY, USA},
    volume = {50},
    number = {6},
    issn = {0360-0300},
    journal = {ACM Comput. Surv.},
    articleno = {78},
    numpages = {37},
}

@article{lewis2020,
    title={Retrieval-augmented generation for knowledge-intensive NLP tasks},
    author={Lewis, P. and Perez, E. and Piktus, A. and Petroni, F. and Karpukhin, V. and Goyal, N. and K{\"u}ttler, H. and Lewis, M. and Yih, W. and Rockt{\"a}schel, T. and others},
    journal={Advances in Neural Information Processing Systems},
    volume={33},
    pages={9459--9474},
    year={2020}
}

@article{katsumi2018,
    title = {Ontologies for transportation research: A survey},
    journal = {Transportation Research Part C: Emerging Technologies},
    volume = {89},
    pages = {53-82},
    year = {2018},
    issn = {0968-090X},
    author = {Megan K. and Mark F.},
}

@article{tupayachi2024,
    author = {Tupayachi, J. and Xu, H. and Omitaomu, O. A. and Camur, M. C. and Sharmin, A. and Li, X.},
    title = {Towards next-generation urban decision support systems through AI-powered construction of scientific ontology using large language models—a case in optimizing intermodal freight transportation},
    journal = {Smart Cities},
    volume = {7},
    year = {2024},
    number = {5},
    pages = {2392--2421},
    issn = {2624-6511},
}

@article{ye2025,
    author = {Ye, S. and Wu, Q. and Fan, P. and Fan, Q.},
    title = {A survey on semantic communications in internet of vehicles},
    journal = {Entropy},
    volume = {27},
    year = {2025},
    number = {4},
    article-number = {445},
    pubmedid = {40282680},
    issn = {1099-4300},
}

@article{zhang2025,
    author={Zhang, Q. and Ma, Z. and Zhang, P. et al.},
    title={Mobility Knowledge Graph: Review and ITS Application in Public Transport},
    journal={Transportation},
    number={52},
    pages={1119–1145},
    year={2025},
}

@inproceedings{lu2025,
    author = {Lu, Y. and Yao, B. and Gu, H. and Huang, J. and Wang, Z. J. and Li, Y. and Gesi, J. and He, Q. and Li, T. J. and Wang, D.},
    title = {UXAgent: An LLM agent-based usability testing framework for web design},
    year = {2025},
    isbn = {9798400713958},
    publisher = {Association for Computing Machinery},
    address = {New York, NY, USA},
    booktitle = {Proceedings of the Extended Abstracts of the CHI Conference on Human Factors in Computing Systems},
    articleno = {545},
    numpages = {12},
    location = {
    },
    series = {CHI EA '25}
}

@inproceedings{yang2024,
    author={Yang, H. and Siew, M. and Joe-Wong, C.},
    booktitle={2024 IEEE International Workshop on Foundation Models for Cyber-Physical Systems \& Internet of Things (FMSys)}, 
    title={An LLM-based digital twin for optimizing human-in-the loop systems}, 
    year={2024},
    volume={},
    number={},
    pages={26-31},
}

@article{masri2025,
    author = {Masri, S. and Ashqar, H. I. and Elhenawy, M.},
    title = {Large language models (LLMs) as traffic control systems at urban intersections: A new paradigm},
    journal = {Vehicles},
    volume = {7},
    year = {2025},
    number = {1},
    article-number = {11},
    issn = {2624-8921},
}

@inproceedings{cui2024,
    author    = {Cui, C. and Ma, Y. and Cao, X. and Ye, W. and Zhou, Y. and Liang, K. and Chen, J. and Lu, J. and Yang, Z. and Liao, K. and Gao, T. and Li, E. and Tang, K. and Cao, Z. and Zhou, T. and Liu, A. and Yan, X. and Mei, S. and Cao, J. and Wang, Z. and Zheng, C.},
    title     = {A survey on multimodal large language models for autonomous driving},
    booktitle = {Proceedings of the IEEE/CVF Winter Conference on Applications of Computer Vision (WACV) Workshops},
    month     = {January},
    year      = {2024},
    pages     = {958-979}
}

@article{nie2025,
    title = {Exploring the roles of large language models in reshaping transportation systems: A survey, framework, and roadmap},
    journal = {Artificial Intelligence for Transportation},
    volume = {1},
    pages = {100003},
    year = {2025},
    issn = {3050-8606},
    author = {T. Nie and J. Sun and W. Ma},
}

@inproceedings{lopez2018,
    author={Lopez, P. A. and Behrisch, M. and Bieker-Walz, L. and Erdmann, J. and Flötteröd, Y. and Hilbrich, R. and Lücken, L. and Rummel, J. and Wagner, P. and Wiessner, E.},
    booktitle={2018 21st International Conference on Intelligent Transportation Systems (ITSC)}, 
    title={Microscopic traffic simulation using sumo}, 
    year={2018},
    volume={},
    number={},
    pages={2575-2582},
}

@article{polak2024,
    author={Polak, M.P. and Morgan, D.},
    title={Extracting accurate materials data from research papers with conversational language models and prompt engineering},
    journal={Nat Commun},
    volume={15},
    number={1569},
    year={2024},
}

@inproceedings{shankar2024,
    author = {Shankar, S. and Zamfirescu-Pereira, J.D. and Hartmann, B. and Parameswaran, A. and Arawjo, I.},
    title = {Who validates the validators? Aligning LLM-assisted evaluation of LLM outputs with human preferences},
    year = {2024},
    isbn = {9798400706288},
    publisher = {Association for Computing Machinery},
    address = {New York, NY, USA},
    booktitle = {Proceedings of the 37th Annual ACM Symposium on User Interface Software and Technology},
    articleno = {131},
    numpages = {14},
    location = {Pittsburgh, PA, USA},
    series = {UIST '24}
}

@article{wood2025,
    author={Wood, J. and Schalkwyk, I.},
    title={Reproducibility in transportation research: Importance, best practices, and dealing with protected and sensitive data},
    journal={Journal of Transportation Technologies},
    volume={15},
    pages={179-202},
    year={2025},
}

@article{gruber1993,
    title = {A translation approach to portable ontology specifications},
    journal = {Knowledge Acquisition},
    volume = {5},
    number = {2},
    pages = {199-220},
    year = {1993},
    issn = {1042-8143},
    author = {T. R. Gruber},
}

@article{studer1998,
    title = {Knowledge engineering: Principles and methods},
    journal = {Data \& Knowledge Engineering},
    volume = {25},
    number = {1},
    pages = {161-197},
    year = {1998},
    issn = {0169-023X},
    author = {R. Studer and V.R. Benjamins and D. Fensel},
}

@article{hogan2021,
    title={Knowledge graphs},
    author={Hogan, A. and Blomqvist, E. and Cochez, M. and d’Amato, C. and Melo, G. D. and Gutierrez, C. and Kirrane, S. and Gayo, J. E. L. and Navigli, R. and Neumaier, S. and others},
    journal={ACM Computing Surveys (Csur)},
    volume={54},
    number={4},
    pages={1--37},
    year={2021},
    publisher={ACM New York, NY, USA}
}

@article{fernandez2016,
    title={Ontology-based architecture for intelligent transportation systems using a traffic sensor network},
    author={Fernandez, S. and Hadfi, R. and Ito, T. and Marsa-Maestre, I. and Velasco, J. R},
    journal={Sensors},
    volume={16},
    number={8},
    pages={1287},
    year={2016},
    publisher={MDPI}
}

@article{mora2022,
    title={Development methodologies for ontology-based knowledge management systems: A review},
    author={Mora, M. and Wang, F. and G{\'o}mez, J. M. and Phillips-Wren, G.},
    journal={Expert Systems},
    volume={39},
    number={2},
    pages={e12851},
    year={2022},
    publisher={Wiley Online Library}
}

@article{janowicz2022,
    title={Know, Know Where, KnowWhereGraph: A densely connected, cross-domain knowledge graph and geo-enrichment service stack for applications in environmental intelligence},
    author={Janowicz, K. and Hitzler, P. and Li, W. and Rehberger, D. and Schildhauer, M. and Zhu, R. and Shimizu, C. and Fisher, C. and Cai, L. and Mai, G. and others},
    journal={AI Magazine},
    volume={43},
    number={1},
    pages={30--39},
    year={2022}
}

@article{gan2025,
    title = {Ontology-driven knowledge graph for decision-making in resilience enhancement of underground structures: Framework and application},
    journal = {Tunnelling and Underground Space Technology},
    volume = {163},
    pages = {106739},
    year = {2025},
    issn = {0886-7798},
    author = {B. Gan and D. Zhang and Z. Huang and F. Zheng and R. Zhu and W. Zhang},
}

@Inbook{McBride2004,
    author="McBride, B.",
    editor="Staab, Steffen
    and Studer, Rudi",
    title="The resource description framework (RDF) and its vocabulary description language RDFS",
    bookTitle="Handbook on Ontologies",
    year="2004",
    publisher="Springer Berlin Heidelberg",
    address="Berlin, Heidelberg",
    pages="51--65",
    isbn="978-3-540-24750-0",
}

@article{moradi2013,
    title = {Knowledge-collector agents: Applying intelligent agents in marketing decisions with knowledge management approach},
    journal = {Knowledge-Based Systems},
    volume = {52},
    pages = {181-193},
    year = {2013},
    issn = {0950-7051},
    author = {M. Moradi and A. Aghaie and M. Hosseini},
}

@article{bubeck2023,
    title={Sparks of artificial general intelligence: Early experiments with gpt-4},
    author={Bubeck, S. and Chandrasekaran, V. and Eldan, R. and Gehrke, J. and Horvitz, E. and Kamar, E. and Lee, P. and Lee, Y. T. and Li, Y. and Lundberg, S. and others},
    journal={arXiv preprint arXiv:2303.12712},
    year={2023}
}

@article{huang2024,
    title = {Dependency-uware neural topic model},
    journal = {Information Processing \& Management},
    volume = {61},
    number = {1},
    pages = {103530},
    year = {2024},
    issn = {0306-4573},
    author = {H. Huang and Y. Tang and X. Shi and X. Mao},
}

@article{kleinsteuber2024,
    title = {Managing provenance data in knowledge graph management platforms},
    journal = {Datenbank Spektrum},
    volume = {24},
    pages = {43-52},
    year = {2024},
    author = {Kleinsteuber, E. and Al Mustafa, T. and Zander, F. and König-Ries, B. and Babalou, S.}
}

@article{bai2025,
    author = {Bai, X. and He, S. and Li, Y. and Yabo, X. and Xin, Z. and Wenli, D. and Jian-Rong, L.},
    title = {Construction of a knowledge graph for framework material enabled by large language models and ITS application},
    journal = {npj Computational Materials},
    volume = {11},
    number = {51},
    year = {2025},
}

@article{gupta2024,
    title={A comprehensive survey of retrieval-augmented generation (RAG): Evolution, current landscape and future directions},
    author={Gupta, S. and Ranjan, R and Singh, S. N.},
    journal={arXiv preprint arXiv:2410.12837},
    year={2024}
}
\end{document}